\documentclass[sigconf]{acmart}

\usepackage[utf8]{inputenc}
\usepackage{graphicx}
\usepackage{subcaption}
\usepackage{enumitem}
\usepackage{kotex}
\usepackage{array,booktabs,tabularx}
\newcolumntype{L}[1]{>{\raggedright\arraybackslash}p{#1}} 
\newcolumntype{Y}{>{\raggedright\arraybackslash}X}        
\AtBeginDocument{}
\usepackage{placeins}
\copyrightyear{2026}
\acmYear{2026}
\setcopyright{cc}
\setcctype{by}
\acmConference[DIS Companion '26]{Designing Interactive Systems Conference}{June 13--17, 2026}{Singapore, Singapore}
\acmBooktitle{Designing Interactive Systems Conference (DIS Companion '26), June 13--17, 2026, Singapore, Singapore}
\acmDOI{10.1145/3802974.3809413}
\acmISBN{979-8-4007-2632-3/2026/06}

\begin{document}

\title{\textbf{Designing Transparent AI-Mediated Language Support for Intergenerational Family Communication}}

\author{Sora Kang}
\email{sorakang@snu.ac.kr}
\affiliation{%
  \institution{Seoul National University}
  \city{Seoul}
  \country{Republic of Korea}
}

\author{Youjin Hwang}
\email{youjin.h@snu.ac.kr}
\affiliation{%
  \institution{Seoul National University}
  \city{Seoul}
  \country{Republic of Korea}
}

\author{Joonhwan Lee}
\email{joonhwan@snu.ac.kr}
\affiliation{%
  \institution{Seoul National University}
  \city{Seoul}
  \country{Republic of Korea}
}

\renewcommand{\shortauthors}{Kang, et al.} 



\begin{teaserfigure}
    \centering
    \includegraphics[width=\linewidth]{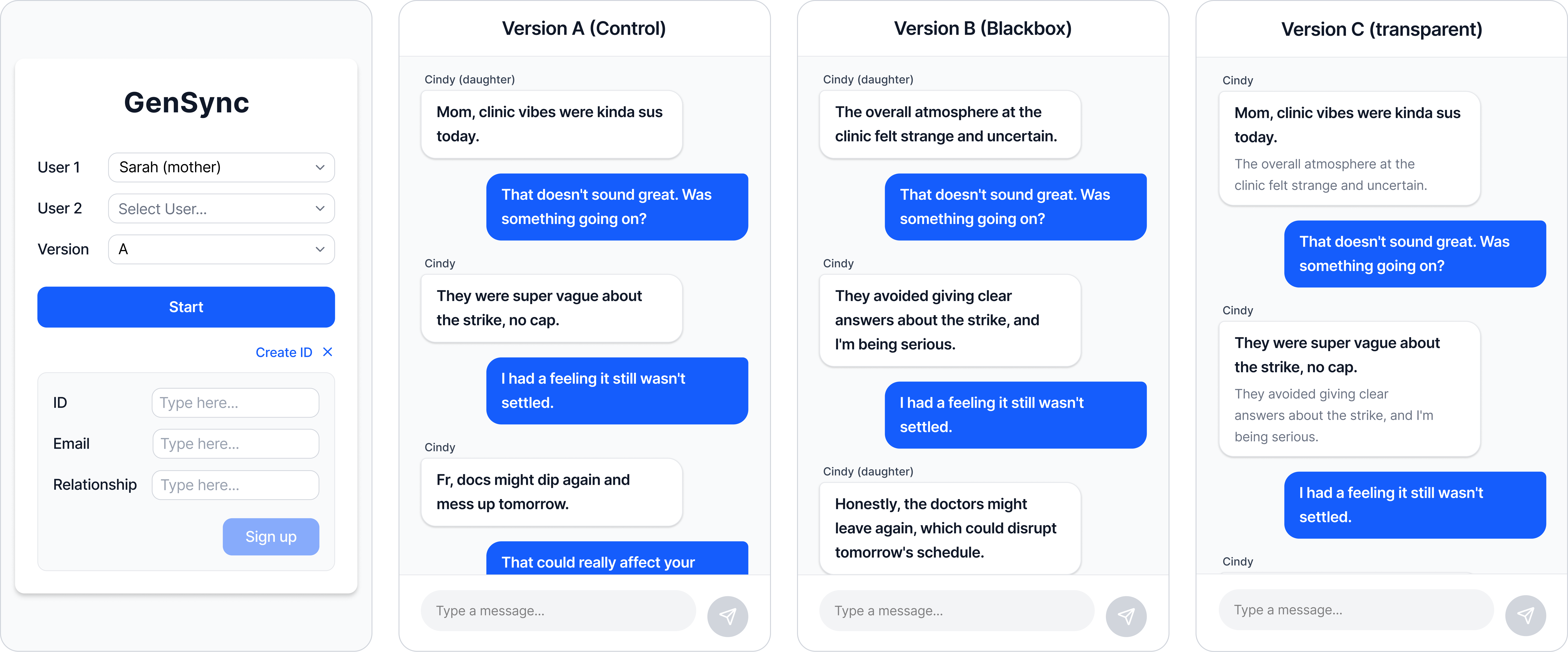}
    \Description{GenSync interface. We evaluated three interface versions: (A) Control, with no translation support; (B) Black-box GenSync, showing only the interpreted message; and (C) Transparent GenSync, showing both the original and interpreted messages.}
    \caption{GenSync interface. We evaluated three interface versions: (A) Control, with no translation support; (B) Black-box GenSync, showing only the interpreted message; and (C) Transparent GenSync, showing both the original and interpreted messages.}
    \label{fig:gensync-interface}
\end{teaserfigure}

\begin{abstract}
Intergenerational linguistic differences pose challenges to effective and intimate family communication. This paper presents GenSync, a chat-based interface that supports intergenerational understanding through different forms of translation visibility. We conducted a controlled within-subjects study with 16 family dyads (32 participants), comparing three conditions: no translation, black-box translation, and transparent translation that displays both original and interpreted messages. The results show that translation visibility plays a critical role in shaping conversational experiences. Transparent translation supported conversational quality, intimacy, and usability, while black-box translation often disrupted conversational flow. These findings position intergenerational language support as a form of interpretive mediation and contribute design implications for AI-mediated communication in socially sensitive contexts.
\end{abstract}

\maketitle

\section{Introduction}
Communication is foundational to human relationships: it enables understanding, empathy, and connection \cite{Adigwe2016,Stiff1988} and supports family well-being and mutual respect \cite{Wang2015}. Beyond exchanging information, family communication helps build emotional bonds \cite{Segrin2018} and trust \cite{Romer2004}. However, contemporary family interactions increasingly face intergenerational linguistic divides, where generation-specific slang, idioms, and online vernacular can make messages hard to interpret across age groups \cite{Shabaniminaabad2020,Clark2009,Barrie2019,Gee2017}. Such gaps may lead to misinterpretations and feelings of distance within families \cite{Gee2017,Onyeator2019}. For example, parents born in the 1970s (and even more so grandparents) may struggle with messages like: “이번 기말고사 완전 억까 에바야 \includegraphics[height=1em]{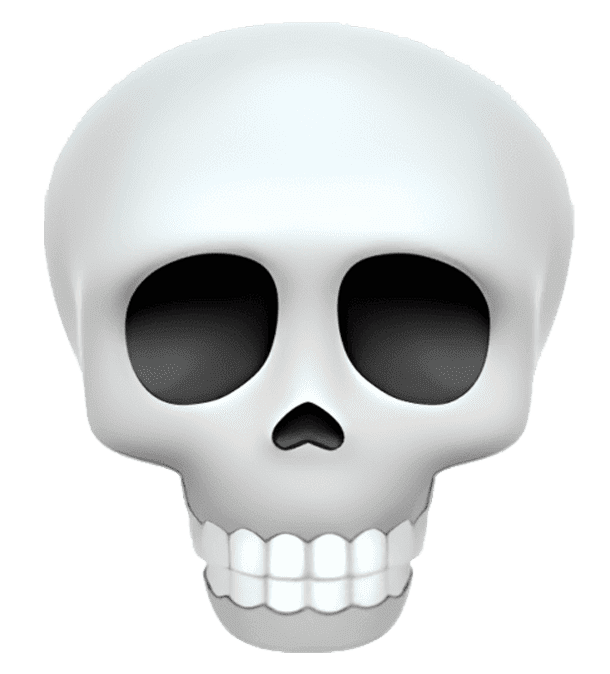}\includegraphics[height=1em]{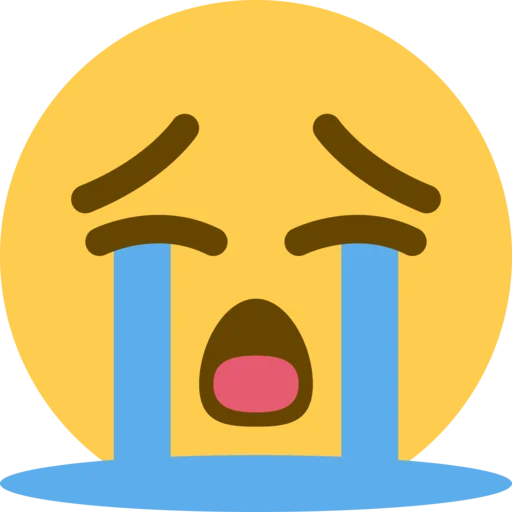}” (ibeon gimalgosa wan-jeon eok-kka eba; “this final exam is completely unfair and crossing the line”).

Recent advances in large language models have renewed interest in using AI to support human communication. Such interventions fall under the domain of AI-Mediated Communication, where an AI's autonomy and role orientation fundamentally shape interpersonal dynamics \cite{Hancock2020}. In socially sensitive contexts like family dialogue, it is crucial to understand how the magnitude of AI intervention—specifically, the visibility of its translations—affects users' sense of agency. However, it remains an open question how these systems should optimally intervene in intergenerational conversations, and how varying forms of intervention shape relational experiences. While people often adjust their vocabulary depending on their audience, intergenerational family interactions introduce additional social and relational constraints that complicate straightforward language adaptation.

To explore this design space, we developed GenSync, a chat-based interface that supports intergenerational understanding by mediating generation-specific expressions. To evaluate GenSync, we conducted a controlled within-subjects study with 16 family dyads (32 participants). Each dyad experienced three interface conditions—(A) Control (no translation), (B) Black-box GenSync (interpreted message only), and (C) Transparent GenSync (original and interpreted messages)—in randomized order. During each condition, dyads engaged in text-based conversations across four everyday contexts (casual, emotional, information exchange, and entertainment-related). Our findings indicate that while translation can support intergenerational understanding, the visibility of translation plays a critical role in shaping how participants verify, interpret, and respond to translated messages in family conversations. Through the design of GenSync and a controlled empirical comparison of translation visibility, this work contributes insights into how AI-mediated language interfaces can be designed for socially sensitive communication contexts.

\section{Method}
\subsection{GenSync}
GenSync is a GPT-4–based chat system designed to support intergenerational communication by rephrasing generation-specific expressions into more interpretable forms for another generation. The system relies on representative intergenerational expressions embedded in the system prompt as few-shot guidance \cite{Brown2020}, enabling contextually appropriate interpretations without task-specific model training.

\subsubsection{Prompt Design}
Each system prompt comprised three components: (1) a task instruction specifying the direction of intergenerational interpretation (e.g., slang to standard language or vice versa), (2) representative slang expressions provided as examples to guide model behavior, and (3) an output specification constraining responses to be concise and suitable for chat-based interaction while remaining sensitive to relational context (e.g., parent–child communication).

\subsubsection{Slang Dictionary}
We constructed a slang dictionary focused on \textit{Kupsikche} (K-slang), a form of generational slang commonly used by Generation Z in South Korea in informal, text-based communication. Publicly available expressions were collected from Naver and curated by two authors to remove obsolete, overly niche, or poorly contextualized items. Each entry included the slang expression, its interpretation in standard Korean, and a brief usage example. The resulting dictionary was incorporated directly into the system prompts as few-shot examples.

\subsubsection{Implementation}
GenSync was implemented as a real-time, chat-based interface following interaction patterns familiar from everyday messaging platforms. For the experiment, we deployed three interface versions differing in interpretive transparency (Figure~1): Version A (Control) provided no interpretation support, Version B (Black-box) displayed only the system-generated interpretation, and Version C (Transparent) presented both the original message and its interpretation.

\subsection{User Study} 
We recruited 16 Korean family dyads (32 participants), each consisting of one younger family member and one older family member. Younger participants were adolescents aged 11–19 who reported frequent use of generation-specific slang in daily communication, while older participants ranged from 22 to 52 years. Dyads included 13 parent–child pairs and 3 sibling pairs. Age gaps between dyad members ranged from 10 to 36 years, and all participants used messaging applications regularly and communicated with their family member at least once per week.
The study was conducted remotely via Zoom, with dyad members joining from separate rooms to simulate everyday text-based communication. After completing a pre-survey, participants were introduced to the GenSync interface and given time to familiarize themselves with its functionality. They were asked to communicate as naturally as they would with their family member. Each dyad experienced all three interface conditions in a randomized order and engaged in conversations across four themes: casual, emotional, informational, and entertainment-related. All interactions were conducted in Korean and subsequently translated into English for analysis by bilingual researchers.

\subsection{Measures and Analysis}
Participants completed surveys during and after the experiment to assess conversation quality, family and intergenerational intimacy, and usability. To assess the system’s impact while minimizing participant fatigue---given our repeated-measures design across four conversation contexts---we selected and adapted single-item measures for each core construct from validated scales. Specifically, perceived usefulness was measured using an item adapted from the Technology Acceptance Model (TAM) \cite{Davis1989}. Conversation quality and relational intimacy were assessed using items derived from established computer-mediated communication (CMC) and family interaction metrics \cite{Wang2015,Segrin2018}. Overall continuance intention was measured based on the Expectation-Confirmation Model \cite{Bhattacherjee2001}. All items were rated on a 5-point Likert scale (1 = Strongly Disagree, 5 = Strongly Agree). 

Quantitative differences across interface conditions were examined using a one-factor repeated-measures ANOVA ($\alpha = .05$) with Bonferroni-corrected paired t-tests for post-hoc comparisons, reporting partial eta squared ($\eta_p^2$) and Cohen’s $d_z$ as effect sizes. Friedman tests (Kendall’s $W$) were additionally conducted as a non-parametric robustness check for the Likert-scale data. Post-study interviews lasting 30--40 minutes were audio-recorded, transcribed, and analyzed using ATLAS.ti \cite{ATLAS} through open coding. Codes were iteratively grouped into themes related to conversation quality, relationship dynamics, and user experience with GenSync to contextualize the quantitative results.

\section{Result}
We report findings from a controlled study comparing three interface conditions: (A) Control (no translation support), (B) Black-box GenSync (interpreted message only), and (C) Transparent GenSync (original + interpreted message). Across conditions, participants rated conversation quality, family intimacy, intergenerational intimacy, and usability on 5-point Likert scales. We use interviews to explain why certain conditions succeeded or failed. We use anonymized participant identifiers (P1–P32) for qualitative excerpts.

\begin{figure}[htbp]
    \centering
    \includegraphics[width=1\linewidth]{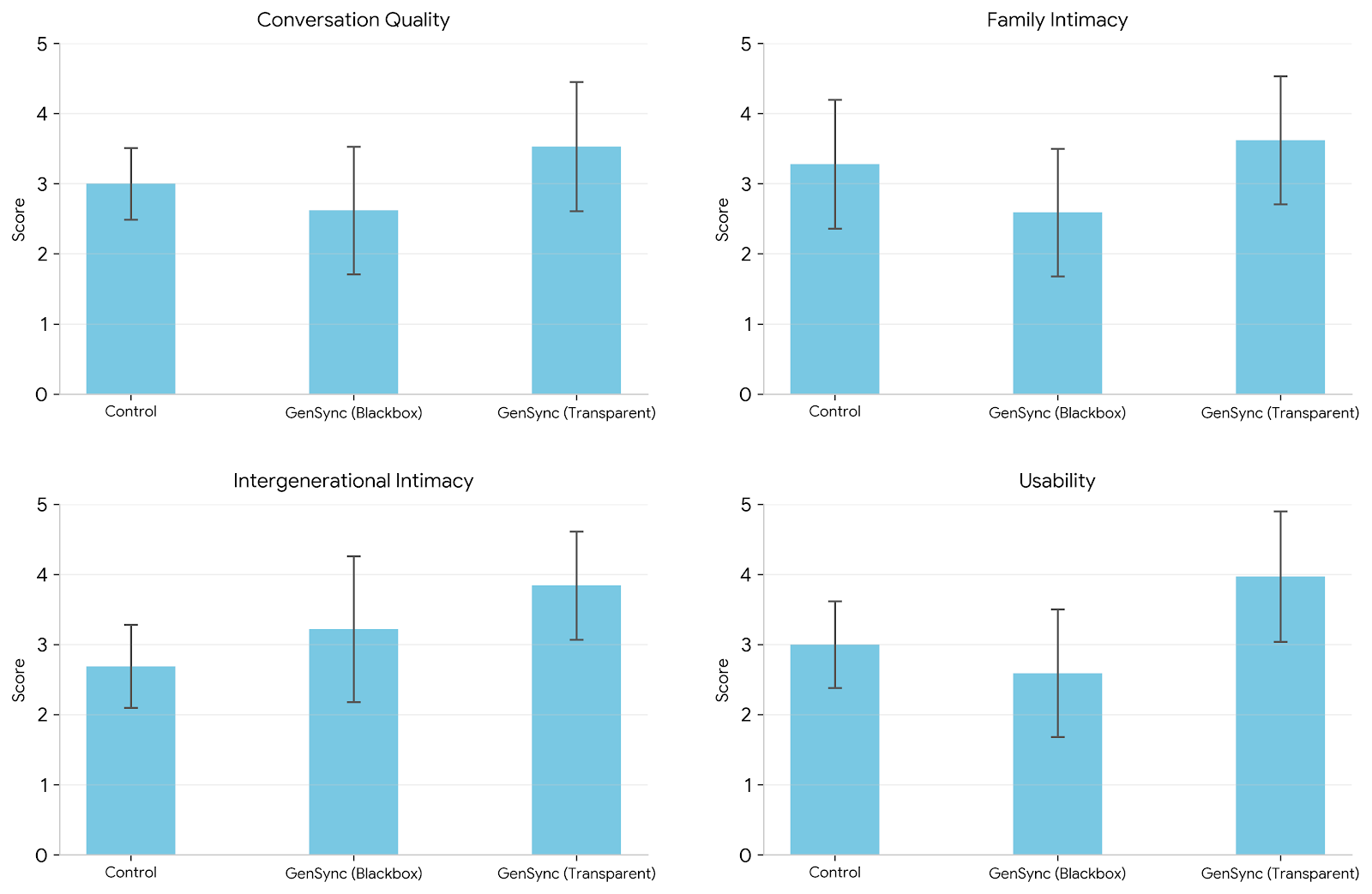}
    \caption{Overall user experience with GenSync. Error bars indicate standard deviations (SD). Transparent mode significantly outperformed both Control and Blackbox conditions across most metrics ($p < .05$).}
    \label{fig:placeholder}
\end{figure}

\subsection{Transparent GenSync improves conversation quality, while black-box translation can reduce it}
Conversation quality differed significantly across conditions ($F(2, 62) = 11.00$, $p < .001$, $\eta_p^2 = 0.262$; Friedman $\chi^2 = 12.87$, $p = .002$, $W = 0.201$). Transparent GenSync (C) achieved the highest ratings ($M=3.53$, $SD=0.92$), followed by Control (A) ($M=3.00$, $SD=0.51$), while Black-box GenSync (B) scored lowest ($M=2.62$, $SD=0.91$). Bonferroni-corrected post-hoc tests showed significant differences for A vs C ($p = .018$, $d_z = -0.52$) and B vs C ($p < .001$, $d_z = -0.77$), but not for A vs B. Interview data suggest that participants valued (C) because seeing both the original and interpreted messages supported topic expansion and conversational flow (e.g., P3, P13, P19). In contrast, participants reported that (B) increased the risk of misunderstanding when inputs contained typos or when the interpretation felt off, because they could not verify what was actually said (e.g., P8, P10, P11, P18, P31). Several participants described this as creating frustration and relational “disconnect” when responses seemed mismatched to intent (e.g., P11, P18).

\subsection{Intimacy benefits depend on what “intimacy” measures}
Family intimacy differed significantly across conditions ($F(2, 62) = 15.05$, $p < .001$, $\eta_p^2 = 0.327$; Friedman $\chi^2 = 26.91$, $p < .001$, $W = 0.421$). Ratings were highest for (C) ($M=3.62$, $SD=0.91$), followed by (A) ($M=3.28$, $SD=0.92$), with (B) lowest ($M=2.59$, $SD=0.91$). Bonferroni post-hoc tests indicated a significant difference for A vs B ($p = .006$, $d_z = 0.60$) and B vs C ($p < .001$, $d_z = -1.20$), but not for A vs C. 

Intergenerational intimacy also differed significantly ($F(2, 62) = 34.56$, $p < .001$, $\eta_p^2 = 0.527$; Friedman $\chi^2 = 42.15$, $p < .001$, $W = 0.659$), but followed a different pattern: (C) remained highest ($M=3.84$, $SD=0.77$), while (B) ($M=3.22$, $SD=1.04$) exceeded (A) ($M=2.69$, $SD=0.59$). Post-hoc tests showed significant differences across all pairs: A vs B ($p = .005$, $d_z = -0.60$), A vs C ($p < .001$, $d_z = -2.25$), and B vs C ($p = .001$, $d_z = -0.69$). Interviews suggest that translation can reduce perceived distance between generations, but relationship outcomes were sensitive to how translations affected perceived “voice” and tone. Some older adults noted that overly formal outputs (especially in black-box mode) made family members sound unlike themselves, reducing genuineness (e.g., P11, P12). By contrast, transparent mode helped participants appreciate each other’s natural styles while still providing interpretive support, which several participants linked to increased understanding and connection (e.g., P1, P17, P21–P22, P30).

\subsection{Usability is strongly tied to visibility and user verification}
Usability differed significantly across conditions ($F(2, 62) = 37.99$, $p < .001$, $\eta_p^2 = 0.551$; Friedman $\chi^2 = 35.64$, $p < .001$, $W = 0.557$). (C) scored highest ($M=3.97$, $SD=0.93$), followed by (A) ($M=3.00$, $SD=0.62$), and (B) ($M=2.59$, $SD=0.91$). Bonferroni post-hoc tests showed significant differences for A vs C ($p < .001$, $d_z = -1.24$) and B vs C ($p < .001$, $d_z = -1.41$), but not for A vs B. Participants described transparent mode as enabling verification, reducing miscommunication risk, and supporting self-directed learning of intergenerational expressions (e.g., P1, P3–P4, P6, P13, P29). At the same time, some teenagers reported that generated slang sometimes felt patterned or misaligned with their personal style, pointing to a need for personalization (e.g., P2, P14, P22).
\FloatBarrier

\section{Design Implications for Intergenerational Language Interfaces}
\subsection{Supporting Interpretive Work Through Interface Visibility}

In our study, black-box translation presented only an interpreted result, and this design choice inadvertently stripped the interaction of transparency, a factor previously identified as critical for maintaining interpersonal trust in AI-mediated communication~\cite{rai2024understanding, papenmeier2019trust}. Consistent with findings by Rai et al.~\cite{rai2024understanding} and Papenmeier~\cite{papenmeier2019trust}, the lack of visibility into the transformation process led to hesitation and reduced engagement. Participants reported that without the original input, they were unable to verify whether awkward phrasing stemmed from a mistranslation or the sender’s actual intent---a limitation that hinders trust calibration~\cite{schmidt2020transparency}. In contrast, transparent GenSync displayed the original message alongside the AI-generated interpretation. This configuration served as explanatory evidence, enabling participants to compare readings, recognize mismatches, and collaboratively repair meaning. By exposing the source text, the interface allowed users to differentiate between linguistic errors and natural language differences, thereby shifting the user's role from a passive recipient to an active verifier. Rather than eliminating uncertainty, transparency supported participants in managing it, transforming the AI from a black box into a referenced resource for maintaining conversational flow.

Together, these observations highlight how interface visibility can shift AI-mediated translation from a mechanism that delivers results to one that supports interpretive work. By preserving access to original expressions and making alternative readings available, interfaces can afford users opportunities to inspect, question, and intervene in meaning-making, particularly in asymmetric communication contexts such as intergenerational exchange.

\subsection{Representational Tensions in AI-Mediated Generational Language}
Beyond issues of interpretive accuracy, our findings reveal tensions related to how users’ language is represented through AI-mediated translation. Several younger participants expressed discomfort when translations altered their tone—either by making older family members appear to “speak like them” or by flattening emotional nuances into standardized formats, which made the interaction feel impersonal or robotic. Notably, these reactions did not stem from misunderstandings of content, but from a sense that the translated output no longer felt like their own way of speaking. In black-box translation, the absence of the original message often made it difficult for participants to distinguish between what had been said and how the system had re-authored it. As a result, some participants described the AI as “speaking for them,” leading to unease about how their language was being conveyed to others. In contrast, transparent GenSync partially mitigated this tension by preserving original expressions alongside interpretive support. This allowed participants to maintain visibility into how their language was transformed, while keeping the AI’s interpretation positioned as an auxiliary resource rather than a replacement for their voice.

These findings point to representational tensions that arise when AI systems intervene in generationally marked language. Rather than aiming to align different generations through stylistic convergence, interfaces that preserve original expressions while offering interpretive support can enable understanding without obscuring differences in how people speak. Such designs foreground questions of representation and voice in AI-mediated communication, emphasizing how language is shown and attributed, not only how it is translated.

\section{Conclusion}
Intergenerational language differences pose persistent challenges for family communication, not just because they are difficult to translate accurately, but because they involve identity, interpretation, and relational judgment. In this work, we presented GenSync as a mediating interface that supports intergenerational understanding by making interpretations visible. Through a mixed-methods study with family dyads, we showed that transparent AI-mediated communication can improve conversation quality, intimacy, and usability, while black-box translation may inadvertently undermine trust and conversational flow. By allowing users to compare original and interpreted messages, GenSync positioned AI as an interpretive aid rather than an authoritative translator, supporting meaning negotiation instead of enforcing convergence toward a shared style. This work does not claim to solve intergenerational language gaps. Instead, it offers a provocation for designers and researchers: to reconsider when AI should translate, when it should explain, and when it should step back to respect human judgment and generational identities. We acknowledge that our findings may be influenced by the cultural specificity of the Korean family context, which emphasizes generational hierarchy (e.g., Confucian values). Future work should include cross-cultural comparisons and longitudinal, naturalistic deployments in everyday messaging applications to observe how these transparent mediations affect relationship dynamics over time. We invite future work to explore how such mediating interfaces might be adapted to other socially sensitive contexts where ambiguity, identity, and relationships are at stake.

\bibliographystyle{ACM-Reference-Format}
\bibliography{gensync}

\end{document}